\documentstyle[11pt,epsfig]{article}

\begin{document}
\begin{titlepage}
  \begin{flushright}
    hep-ph/0410302
  \end{flushright}
  \begin{center}
    \vspace*{1.4cm}
    
  {\Large\bf Hierarchical Quark Masses and Small Mixing Angles\\
    from Warped Intersecting Brane Models}\vspace*{1cm}
  
  Tatsuya Noguchi\footnote{E-mail: noguchi@tuhep.phys.tohoku.ac.jp}
  \vspace*{5mm}
  
  $^*${\it Tohoku University, Miyagi 980-8578, Japan}

  \vspace{1.2cm}
  
  \begin{abstract}
    We propose a novel mechanism for reproducing the realistic
    hierarchical structure of the observed CKM mixing matrix and quark
    masses by means of introducing a warped metric.  We illustrate the
    method on the basis of a specific Type IIA intersecting brane
    model proposed by Ib\'a\~nez et al.  The model is compactified on
    a direct product of three tori, on which four D-brane stacks wrap,
    and the intersections between two stacks collectively provide
    three generations of quarks and leptons. The exponential of the
    area formed by the intersection points corresponding to three
    matter fields is interpreted as the Yukawa coupling among them.
    It is known, however, that the Yukawa matrices for quarks are
    generically singular, and that it does not give definite mixing
    angles nor hierarchically suppressed quark masses. We show that
    the newly introduced warp effect on the internal manifold modifies
    the Yukawa matrix elements and generate hierarchical quark masses
    and CKM mixing angles.
\end{abstract}

\end{center}
\end{titlepage}

\setcounter{footnote}{0}

\section{Introduction}
\setcounter{equation}{0}

The string theories have been regarded as one of the most promising
models for describing both gravity and particle physics.
Many researchers have tried to derive the Standard Model within the
framework. Since the theories are originally constructed in higher
dimensional spacetime, the extra dimensions must be compactified on
some manifold. 
Through these procedure, in most cases, there appear many scalar
fields at low energy scale, and their vacuum expectation values remain
unfixed because we cannot calculate the potentials of these fields
within the perturbative framework.

In the mid-1990s, the author in Ref.~\cite{Polchinski:1995mt}
formulated D-branes, spatially extended objects, on the worldsheet of
string theories by imposing Dirichlet boundary conditions on the end
points of strings. It 
is found that D-branes carry a certain amount of Ramond-Ramond charge, and
that they could be regarded as solitons in supergravity theories.  The
emergence of solitons, as is known from quantum field theories,
suggests a duality transformation from one theory
described in terms of a certain set of fundamental fields and
couplings to another in different terms. This string duality shed
light on the relation among string theories, and invoked a
considerable number of studies on the construction of phenomenological
models with D-branes in Type IIA, Type IIB, and Type I string theories
although they are once regarded inadequate for explaining the real
world.

Recently, intersecting D-brane models have been intensively studied.
The scenario was first proposed by Blumenhagen et
al. in the context of Type I string theory
\cite{Blumenhagen:2000wh,Blumenhagen:2000vk,Blumenhagen:2000ea}.
On a stack of $N$ D-branes, 
the charges assigned to the end points of
open strings, called Chan-Paton factors, are known to form
collectively the adjoint representation of $U(N)$ gauge group.
Furthermore, at the intersection between a stack of
$N$ D-branes and that of $M$ D-branes there appears a massless
fermionic field belonging to a bifundamental representation $(N, \bar
M)$ of a gauge group $U(N)\times U(M)$
\cite{Berkooz:1996km,Arfaei:1996rg}. In the framework of 
intersecting brane models, 
these massless open string modes are identified
with matter fields such as quarks and leptons.
It is worth noting that a Yukawa coupling is described as $e^{-kA}$,
where $k$ 
is a constant of order 1, and $A$ is the area formed by three
intersecting points corresponding to the relevant matter fields.
Since the superstring theories are constructed in 10 dimensional
spacetime, we have to compactify them on 6 dimensional manifold
when it comes to discussing phenomenology.
In that case, D-branes wrap on some cycles of the manifold and they
generically intersect with each other several times, and the
intersection numbers determine the number of matter fields.
Many researchers have investigated intersecting brane models in
various aspects: the search for the standard model  
\cite{Aldazabal:2000dg,Aldazabal:2000cn,Ibanez:2001nd,Cremades:2002cs,Cremades:2003qj}
, the construction of grand unified 
theories\cite{Kokorelis:2002ip,Ellis:2002ci,Cvetic:2002pj,Axenides:2003hs}, 
supersymmetric models
\cite{Cvetic:2001tj,Cvetic:2001nr,Cvetic:2002qa,Honecker:2004kb}
and composite models\cite{Kitazawa:2004ed,Kitazawa:2004nf}, the CFT
calculation of couplings between fields locating at
intersections\cite{Cvetic:2003ch,Abel:2003vv,Lust:2004cx}, 
 quark masses and a CKM mixing matrix
\cite{Cremades:2003qj,Kitazawa:2004nf,Cvetic:2002wh,Chamoun:2003pf}, and  
 phenomenological viability\cite{Klebanov:2003my,Abel:2003fk,Abel:2003yh}.

In this paper, we focus on a specific Type IIA intersecting brane
model compactified on a direct product of three tori proposed by
Ib\'a\~nez et al.\cite{Ibanez:2001nd}
In the model, 
the family of right-handed quarks is distinguished by 
the difference of intersecting points in position on the first torus,
and that of left-handed ones on the third torus.
The Yukawa matrices are described by the area formed by three
intersecting points corresponding to quarks and a Higgs boson.
Owing to the direct product structure of the internal manifold, the
Yukawa matrices take the form of $a_i b_j^{U/D}$ and have
only one non-zero eigenvalue, which is identified to be the third
generation quark mass. In addition, we cannot obtain a definite CKM
mixing matrix in the model.
In this paper, we investigate quark masses and a CKM mixing matrix
assuming a warp effect on the internal manifold, i.e., the dependence
of the first torus volume on a coordinate of the third one. 
We reproduce the realistic hierarchical structure of the observed CKM
mixing matrix and quark masses under some condition on small
parameters originated from the warp effect and the size of tori.

The outline of our paper is as follows.
In section \ref{sec:ibms}, we briefly review the construction of
intersecting brane models. In section \ref{sec:warp}, assuming a warp
effect on tori we calculate the Yukawa matrices and obtain the mass
spectra for quarks and the CKM matrix. We show that our model
reproduces the realistic hierarchical structure with several small
parameters determined properly.
The section \ref{sec:discussion} is devoted to the discussion.

\section{Intersecting Brane Models}
\label{sec:ibms}

We briefly review an intersecting brane model presented in
Ref.~\,\cite{Ibanez:2001nd}.
We consider Type IIA string theory compactified on orientifold six
tori, with several stacks of D-branes intersecting each other at angles.
Let us assume that orientifolds are located at $x_4 = 0,
\pi$, $x_6 = 0, \pi$ and $x_8 = 0, \pi$, and that D6-branes wrapped on a direct product of 1-cycles of each
torus $T^i$ with a set of winding numbers $(n_a^i, m_a^i)$.
The intersection is four dimensional.
We know that at the intersection between $N_a$ D6-branes and $N_b$ D6-branes, there
is an $(N_a, \bar N_b)$ representation of the gauge group
$U(N_a)\times U(N_b)$. In our setup, there are also mirror branes
owing to the orientifolds. The winding number of mirror branes are
described as $(n_a^i, -m_a^i)$. 
Later on, two sectors play an important role in generating quarks and
leptons. The first sector $D_a$-$D_b$, realized on the intersection between
two D-brane stacks, D$6_a$-branes and D$6_b$-branes. The intersection
number turns out to be
\begin{eqnarray}
  \label{eq:int1}
    I_{ab} &=& (n_a^1 m_b^1 - n_b^1 m_a^1)(n_a^2 m_b^2 - n_b^2 m_a^2)
(n_a^3 m_b^3 - n_b^3 m_a^3),
\end{eqnarray}
and the sector gives a bifundamental representation $(N_a, \bar N_b)$. 
The second sector $D_a$-$\Omega D_b$ is realized 
on the intersection between $D6_a$-branes and a mirror stack
corresponding to another D-brane stack, $\bar D6_b$-branes. The
intersection number is
  \begin{eqnarray}
  \label{eq:int2}
    I_{ab} &=& -(n_a^1 m_b^1 + n_b^1 m_a^1)(n_a^2 m_b^2 + n_b^2 m_a^2)
(n_a^3 m_b^3 + n_b^3 m_a^3),
  \end{eqnarray}
and it gives a representation $(N_a, N_b)$.
A negative sign in (\ref{eq:int1}) and (\ref{eq:int2}) represents the
opposite chirality.
The other sectors are not relevant to our discussion.

So far, we have not imposed any restriction on winding numbers. 
We have to take account of Ramond-Ramond tadpole cancellation
condition, which means the cancellation of D-brane charge on the internal manifold. If there is no orientifold, the condition is described as
\begin{eqnarray}
  \prod_a^K N_a\Pi_a = 0,
\end{eqnarray}
i.e. the sum of the cycles on which D-branes wrap is equal to zero.
If there are several orientifolds, they play a role of negatively charged
objects. In our setup, the orientifold planes wrapping on a cycle $(1,
0), (1, 0), (1, 0)$, the condition reads
\begin{eqnarray}
\label{eq:tadpole}
  \sum_a N_a n_a^1n_a^2n_a^3 &=& 16,\nonumber\\
  \sum_a N_a m_a^1m_a^2n_a^3 &=& 0,\nonumber\\
  \sum_a N_a m_a^1n_a^2m_a^3 &=& 0,\nonumber\\
  \sum_a N_a n_a^1m_a^2m_a^3 &=& 0.
\end{eqnarray}

We should consider the possibility of NS B-flux $b^i$ on a torus $T^i$
\cite{Blumenhagen:2000ea}.
Starting with Type I string theory compactified on a direct product
of three tori, the introduction of NS B-flux
background $b = 1/2$ on a torus modifies its complex structure. In the intersecting brane picture, the effect is equivalent to
changing a winding number on an original torus $(n, m')$ into $(n,
m'+n/2)$. We will describe the configuration of D-branes in terms of
this newly defined effective winding number $m = m'+n/2$ on a torus
with non-zero $B$-field background.
\footnote{
The existence of orientifolds on tori requires the quantization of the
$B$ flux $b = 0, 1/2$.
}

We are now ready to look for winding numbers generating standard model
matter fields: 
\begin{center}
\label{eq:desirable matters}
  $Q_L^i = (3, 2, 1/6);\quad U_R^i = (\bar 3, 1, -2/3);\quad D_R^i = (\bar 3, 1, 1/3);$\\
$L^i=(1, 2, -1/2);\quad E_R^i = (1, 1, 1)$.
\end{center}
Note that we have to put at least four stacks of D-branes. This is
because the right-handed charged lepton $E_R^i$ does not have the
charge of $SU(3)$ nor that of $SU(2)$, thus at least two more D-brane stacks
are required. From now on, we discuss the models with four D-brane
stacks: stack $a$ composed of 3 D6-branes, stack $b$ of 2 D6-branes, and
the others composed of a D6-brane. 

First, the requirement of the above matter
fields allows only the winding numbers denoted in Table
\ref{tab:winding}, where
 $\beta^i = 1 - b^i$, $\epsilon = \pm 1$ and $\rho = 1, 1/3$.
The variables $n_a^2$, $n_b^1$, $n_c^1$ and $n_d^2$ take integers.
Note that on the third torus, the $B$ field always takes a value
$b^3 = 1/2$. 
This model contains four $U(1)$ charges originated from
  D-brane stacks $Q_a$, $Q_b$, $Q_c$, and $Q_d$. Each of them can be
  understood in terms of the Standard Model framework:
  \begin{eqnarray}
    Q_a = 3 B, \quad
    Q_c = 2 I_R, \quad
    Q_d = -L,
  \end{eqnarray}
where $B$ is the baryon number, $L$ is the lepton number, and $I_R$ is
the third component of right-handed weak isospins. $Q_b$ plays a role
of a generation dependent Peccei-Quinn symmetry. 
Owing to the couplings to the Ramond-Ramond fields, three of the four
$U(1)$ charges become massive and only the gauge boson corresponding
to the hypercharge
\begin{eqnarray}
Q_Y = \frac{1}{6}Q_a - \frac{1}{2}Q_c + \frac{1}{2}Q_d  
\end{eqnarray}
remains massless if and only if the following relation holds: 
\begin{eqnarray}
  \label{eq:condition1}
n_c^1 &=& \frac{\beta^2}{2\beta^1}(n_a^2 + 3\rho n_d^2).
\end{eqnarray}

\begin{table}[htbp]
  \label{tab:winding}
  \centering
  \begin{tabular}[]{|c||c|c|c|}
\hline
  $N_i$ & $(n_i^1,m_i^1)$ & $(n_i^2,m_i^2)$ & $(n_i^3,m_i^3)$\\
\hline\hline
$N_a = 3$ & $(1/\beta^1, 0)$ & $(n_a^2,\epsilon \beta^2)$ &
  $(1/\rho,1/2)$\\ 
\hline
$N_b = 2$ & $(n_b^1,-\epsilon\beta^1)$ & $(1/\beta^2, 0)$ & $(1,
  3\rho/2)$\\ 
\hline
  $N_c = 1$ & $(n_c^1,3\rho\epsilon\beta^1)$ & $(1/\beta^2, 0)$ & $(0,
  1)$\\ 
\hline
  $N_d = 1$ & $(1/\beta^1, 0)$ & $(n_d^2, -\beta^2\epsilon/\rho)$ & 
  $(1, 3\rho/2)$\\
\hline
  \end{tabular}
\caption{The winding numbers yielding the Standard Model
  matter fields.}
\end{table}

Secondly, as for the tadpole cancellation, these winding numbers
satisfy automatically all but the first condition of
Eqs.~(\ref{eq:tadpole}). 
It reads
\begin{eqnarray}
  \label{eq:condition2}
  \frac{3n_a^2}{\rho\beta^1} +   \frac{2n_b^1}{\beta^2} +
  \frac{n_d^2}{\beta^1} = 16.
\end{eqnarray}
If there are more than four D-brane stacks, their contribution should
be added to the left handed side of Eq.~(\ref{eq:condition2}).
It is possible to satisfy tadpole cancellation
conditions by adding some D-brane stacks which do not intersect
with the first four stacks, i.e. keeping the intersection numbers the same.

In conclusion, the desired configuration is realized by the winding numbers
defined in Table \ref{tab:winding} with the  
parameters satisfying Eqs.~(\ref{eq:condition1}) and (\ref{eq:condition2}).
The resultant matter contents and their $U(1)$ charges are shown in Table \ref{tab:matter contents}.
We are now ready to consider the Yukawa couplings.
For simplicity, we take a specific D-brane configuration and calculate
the Yukawa matrices 
\begin{eqnarray}
\beta^1 = \beta^2 = 1,\qquad\quad\;\\
\epsilon = -1,\quad \rho = 1,\qquad\quad\\
a^2 = 2,\; n_b^1 = 0,\; n_c^1 = 1,\; n_d^2 = 0.
\end{eqnarray}
\begin{table}
\begin{center}
  \begin{tabular}[]{|c||c|c|c|}
\hline
  $N_i$ & $(n_i^1,m_i^1)$ & $(n_i^2,m_i^2)$ & $(n_i^3,m_i^3)$\\
\hline\hline
$N_a = 3$ & $(1, 0)$ & $(2, -1)$ & $(1, 1/2)$\\
\hline
$N_b = 2$ & $(0, 1)$ & $(1, 0)$ & $(1, 3/2)$\\
\hline
$N_c = 1$ & $(1, -3)$ & $(1, 0)$ & $(0, 1)$\\
\hline
$N_d = 1$ & $(1, 0)$ & $(0, 1)$ & $(1, 3/2)$\\
\hline
  \end{tabular}
\end{center}
\caption{A combination of winding numbers yielding the SM matter fields}
\label{tab:concrete}
\end{table}

\begin{table}
\begin{center}
  \begin{tabular}[]{|c||c|c|c|c|c|c|c|}
\hline
  Intersection & Matter fields & repr. & $Q_a$ & $Q_b$ & $Q_c$ & $Q_d$ & $Y$\\
\hline\hline
$(a, b)$ & $Q_L$ & $(3, 2)$    & $1$ & $-1$ & $0$ & $0$ & $1/6$\\
\hline
$(a, b^*)$ & $q_L$ & $2(3, 2)$ & $1$ & $1$ & $0$ & $0$ & $1/6$\\
\hline
$(a, c)$ & $U_R$ & $3(\bar 3, 1)$ & $-1$ & $0$ & $1$ & $0$ & $-2/3$\\
\hline
$(a, c^*)$ & $D_R$ & 3$(\bar 3, 1)$ & $-1$ & $0$ & $-1$ & $0$ & $1/3$\\
\hline
$(b, d^*)$ & $L$ & $3(1, 2)$ & $0$ & $-1$ & $0$ & $-1$ & $-1/2$\\
\hline
$(c, d)$ & $E_R$ & $3(1, 1)$ & $0$ & $0$ & $-1$ & $1$ & $1$\\
\hline
$(c, d^*)$ & $N_R$ & $3(1, 1)$ & $0$ & $0$ & $1$ & $1$ & $0$\\
\hline
  \end{tabular}
\end{center}
\caption{Matter contents and $U(1)$ charges}
\label{tab:matter contents}
\end{table}

\begin{figure}[htbp]
\centering
\hspace{-1cm}
\scalebox{.6}{\includegraphics{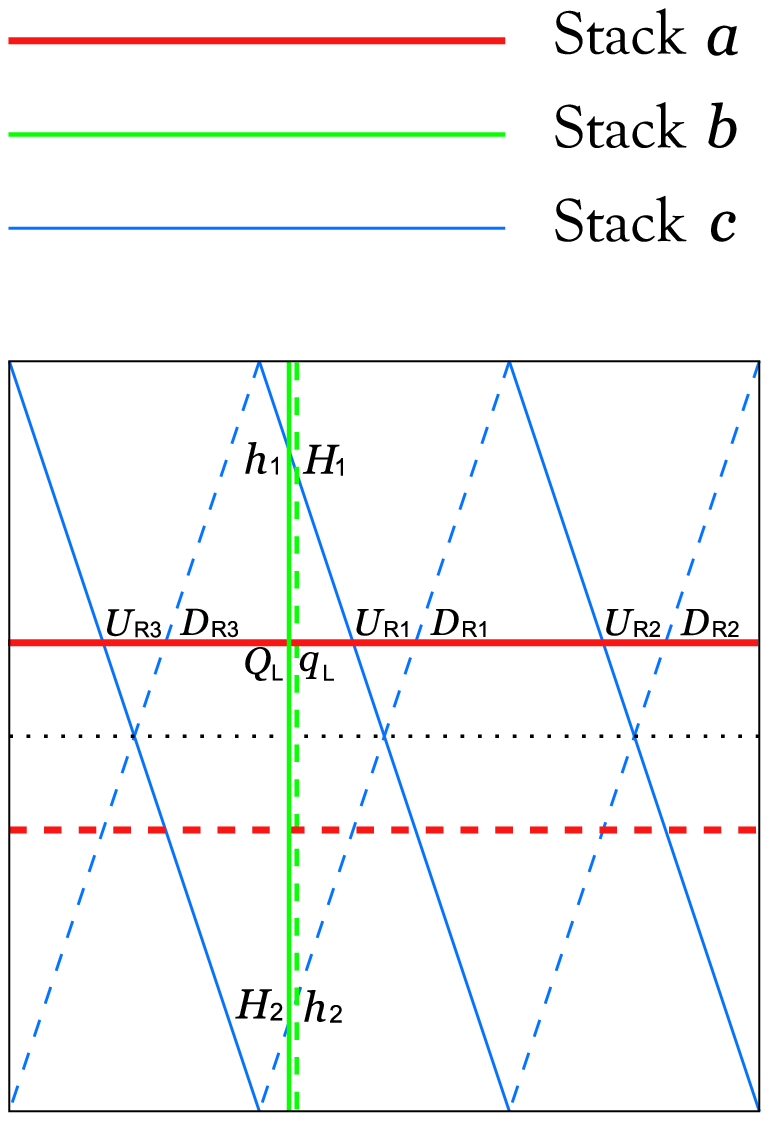}}
\scalebox{.6}{\includegraphics{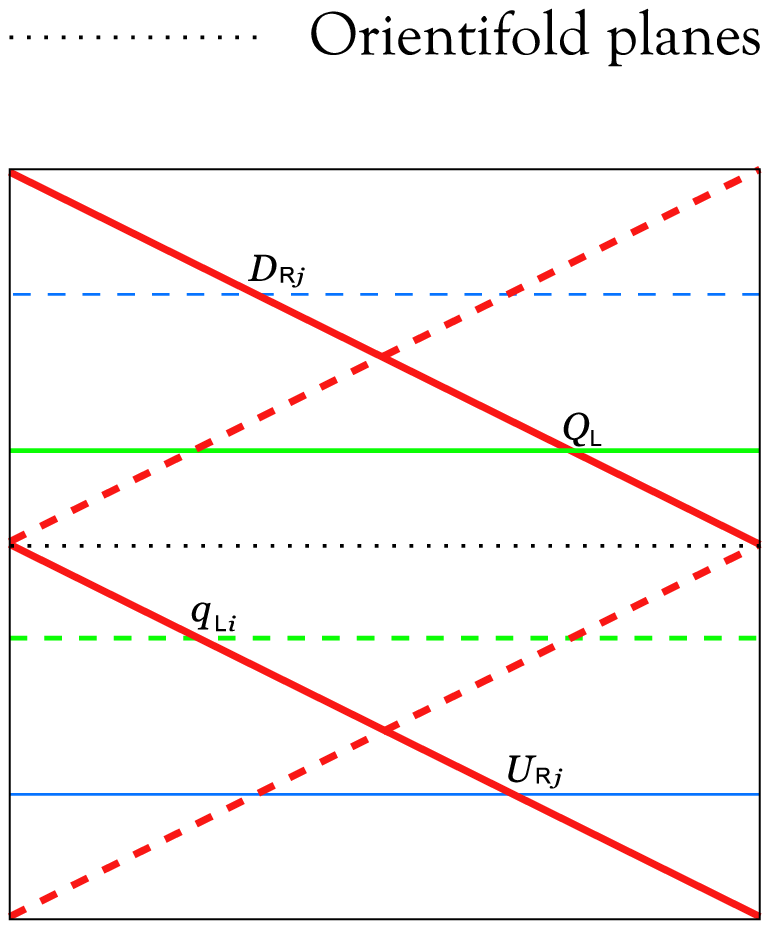}}
\scalebox{.6}{\includegraphics{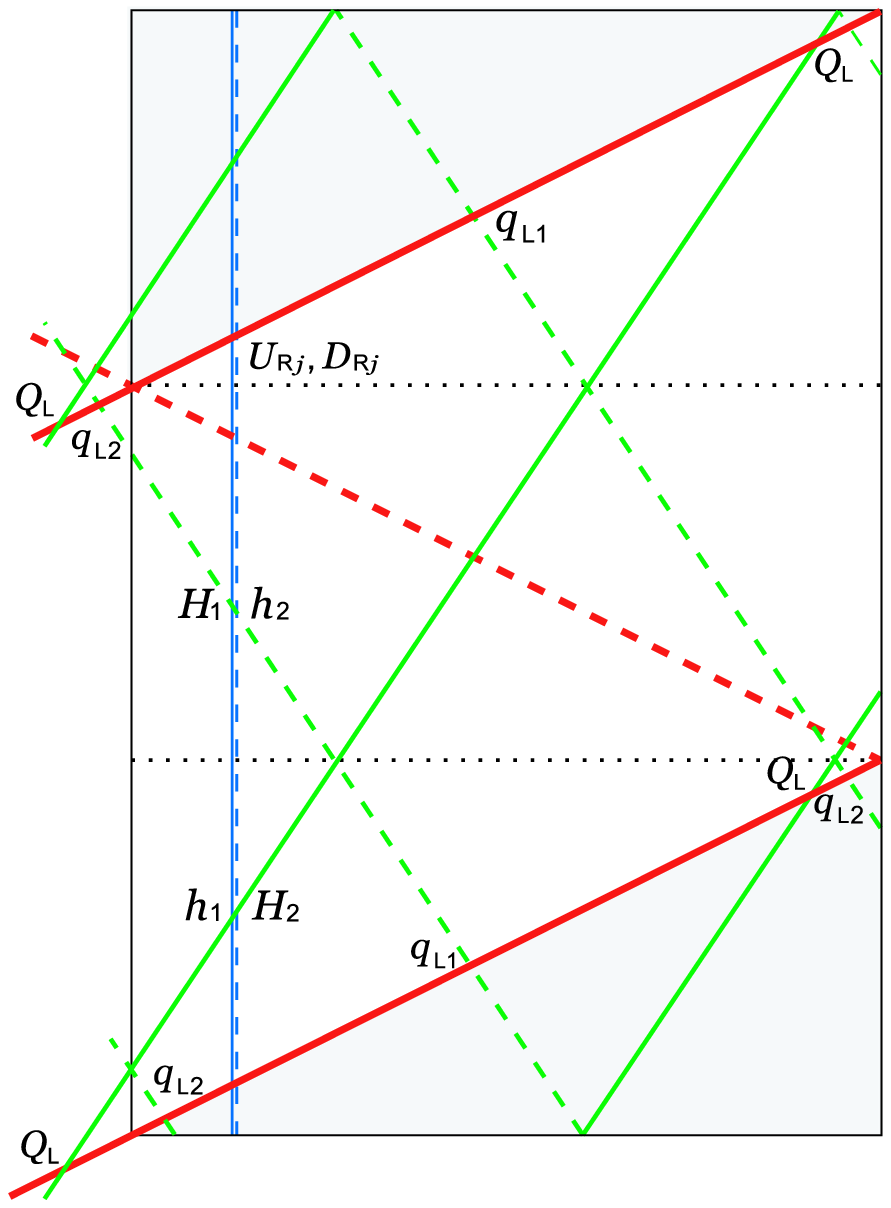}}
% Normal Font -> Georgia
% Subscript -> Arial Regular
% Subscript Number -> Arial Italic
\caption{The configuration of D-brane stacks on the first, second, and
  third torus (from left to right). The stack $a$ is composed of 3
  D6-branes, the stack $b$ of 2 D6-branes, and the stack $c$ is a
  single D6-brane. The broken lines and the dotted ones represent
  mirror branes and orientifold planes, respectively.
Since D$6_d$-branes are not used for
  our analysis on quark sectors, they are not depicted. On the second
  torus, the open strings stretching between $b$-$c^*$ and $b$-$c$
  correspond to the $H_i$'s, and the $h_i$'s. On the third torus, the
  unshaded portion is a fundamental region. 
}
\label{fig:torus123}
\end{figure}

The brane configuration of this setup is depicted in
Fig.~\ref{fig:torus123}, where we take specific values for moduli
parameters indicating positions of branes. 
In order to satisfy the first tadpole cancellation condition, we add
5 parallel D-branes with a winding number $(1,0)\;(1,0)\;(1,0)$,
which do not intersect with any D-brane stacks.

In this context, the Higgs scalar fields come from the open string
modes stretching from two parallel D-brane stacks not from an
intersection. These modes are tachyonic, and are expected to undergo 
tachyon condensation, which could be liken to the Higgs mechanism, 
 and to become finally ordinary Higgs fields with appropriate masses. 
Nevertheless, the mechanism of condensation has not been clarified.
For the purpose of avoiding the complexity on the tachyon
condensation, we assume that the size of the second torus is
so small compared with the others that we neglect it when evaluating
Yukawa couplings. 

This model has the following Yukawa interactions:
 \begin{eqnarray}
 &&  y_j^U Q_LU_R^j h_1 +   y_j^D Q_LD_R^j H_2  \nonumber\\
 &+&  y_{ij}^u q_L^i U_R^j H_1 +   y_{ij}^d q_L^i D_R^j h_2 \nonumber\\
 &+&  y_{ij}^L L^i E_R^j H_2 + y_{ij}^N L^i N_R^j h_1 + \mbox{h.c.}
 \end{eqnarray}
where $Q_L$ and $q_L^i (i = 1,2)$ are left-handed quarks, $U_R^j
(j=1,2,3)$ are right-handed ones, and $h_1$, $h_2$, $H_1$, and $H_2$
are Higgs fields. The coefficients $y^U, y^D, y^u, y^d,
y^L$, and $y^N$ are Yukawa couplings. Each of them is described
by the triangular area $A_{ijk}$ formed by three intersecting points
corresponding to the fields included in the Yukawa interaction term
\begin{eqnarray}
y_{ijk} = e^{-k A_{ijk}},
\end{eqnarray}
where $k$ is a constant of order 1, which is expected to be fixed
through stringy calculation\cite{Cvetic:2003ch,Abel:2003vv}.

This expression can be interpreted intuitively as follows.
All the matter fields such as quarks and leptons come from open string
modes, with one end on a stack and the other end on the other stack.
For example, $Q_L^1$ has one end on D$6_a$-branes and the other on
D$6_b$-branes. Likewise, $U_R^1$ is the open string mode stretching
between D$6_b$-branes and D$6_c$-branes, and $h_1$ between
D$6_c$-branes and D$6_a$-branes. The Yukawa coupling among these three
fields evaluated by the scattering amplitude among these
open strings, i.e., by the exponential of the area swept by them.
When the compact manifold is a direct product of tori, Yukawa
couplings are expressed by the sum of the triangular areas projected
on each torus $A_{ijk}^{(n)}$\cite{Cremades:2003qj}:
\begin{eqnarray}
A_{ijk} = \sum_{n=1}^3 A_{ijk}^{(n)}.
\end{eqnarray}

Before we come to the main subject, we must draw attention to the
stability of intersecting brane models. It is known that, in general,
parallel D-branes do not attract each other because of the
cancellation between the repulsive force coming from RR sectors and
the attractive one from
NS sectors. However, as for D-branes intersecting at angles, this is
not the case.
Besides the massless fermion states, in general, there are some scalar
states at each intersection, which could be regarded as
supersymmetric partners of the fermion states. The mass square of
these scalar states is parameterized by the angles between two
D-brane stacks. Some 
choices of angles give a negative mass square, which suggest the
instability of the configuration. However, with the winding numbers
adopted here, we could choose the radii of tori with which tachyons do
not appear at any intersection. 

\section{The intersecting brane model with a warp effect}
\label{sec:warp}

We now show that a warp effect plays an important role in
generating hierarchically suppressed fermion masses and CKM mixings.  
First, let us consider the up-quark sector. We define
$A_{U,ij}^{(n)}$ as the area of a
triangle formed by three intersection points $Q_{Li}$, $U_{Rj}$ and
$X_{U,i}$ on the $n$-th torus, where $Q_{Li} = q_i\;(i = 1, 2)$, 
$Q_{L3} = Q_L$, $X_{U,i}= h_1\;(i = 1, 2)$ and $X_{U,3} = H_1$.

Then the Yukawa matrix for the up sector is
described as $Y_{U,ij} = \langle X_{U,i} \rangle e^{-kA_{U,ij}}$, where
$A_{U,ij} = A_{U,ij}^{(1)} + A_{U,ij}^{(3)}$. If there is no warping, owing to
the degeneracy of the locations of quark doublets on the first torus,
$A_{U,ij}^{(1)}$ depends only on the
index $j$, where let us call it $A^U_j$. On the other hand, on the third torus the right
handed quarks stay at the same point, thus we define $A^Q_i = A_{U,ij}^{(3)}$.
Therefore, the Yukawa matrix for up-sector quarks is described as
\begin{eqnarray}
\label{eq:aibuj}
  Y_{U,ij} = \langle X_{U,i} \rangle a_i b^U_j
\end{eqnarray}
where $a_i = e^{-kA^Q_i}$ and $b^U_j = e^{-kA^U_j}.$
For the down-quark sector, we obtain 
\begin{eqnarray}
    Y_{D,ij} = \langle X_{D,i} \rangle a_i b^D_j
\end{eqnarray}
where $b^D_j = e^{-kA^D_j}$, and $A^D_j$ is the area formed by $Q_{Li}$,
$D_{Rj}$ and $X_{D,i}$ on the first torus, where $X_{D,i}= h_2\;(i =
1, 2)$ and $X_{D,3} = H_2$. For simplicity, we assume all the
vacuum expectation values of Higgs fields to be unity throughout this paper.

This matrix cannot reproduce the realistic structure of CKM matrix and
quark masses although it seems roughly consistent with them.
Since this matrix is singular, we cannot have
 a definite unitary matrix diagonalizing $Y\cdot Y^\dagger$.
In addition, the resultant CKM mixing
matrix does not have any mixing between the third generation and the
other ones.
Besides, only one of three eigenvalues is non-zero for each sector,
which means that the model cannot explain the masses of quarks and
leptons except for the third generation. 
As for this problem, 
several attempts have been done for generating
definite non-zero mixings.
For example, the authors in Ref.~\cite{Kitazawa:2004ed,
  Kitazawa:2004nf} argue on 
the basis of supersymmetric composite models constructed in the
framework of intersecting brane model. The authors in
Ref.~\cite{Chamoun:2003pf} study the CKM mixing matrix in the model
with 3 supersymmetric Higgs doublets, where all the matter fields
arise from one of the tori.

Let us assume here that the metric of the first torus depends on a
coordinate on the third 
torus $x_8$.  Then, the length of strings projected on the first torus
become dependent on the coordinate $x_8$, and so does the area 
swept by the strings.
In this case, Yukawa couplings cannot be expressed in the simple form
(\ref{eq:aibuj}). 
In a rough approximation, we take the effect of
the $x_8$-dependence into consideration by modifying $A^{(1)}_j$
\begin{eqnarray}
  A^{U}_j\rightarrow A^{U}_{j}(1 - \epsilon_i),
\end{eqnarray}
where $\epsilon_i$ is a small quantity generated by the
 $x_8$-dependence. The $\epsilon_i$ is, in
 principle, calculable in string theory, but we treat them simply as
 input parameters. 
In this estimation, Yukawa couplings are no longer of the
form (\ref{eq:aibuj}), and thus Yukawa matrices would generically
become non-degenerate. Surprisingly, this slight modification gives
 the realistic hierarchical structure of fermion masses and a 
CKM mixing matrix. 
In this paper, we attribute the $x_8$-dependence of the metric to a
warp effect of the background geometry.\footnote{The dependence might be explained by other mechanism.
For example, in the context of M-theory compactified on $S^1/Z_2\times
K_6$, where $K_6$ is a Calabi-Yau manifold, 
the authors in Ref.~\cite{Witten:1996mz} solved the equation of motion
of 11D supergravity by using a strong coupling expansion, and found
that the volume of the Calabi-Yau manifold can linearly depends on the
coordinate of $S^1/Z_2$.}
Before calculating quark masses and CKM mixings,
we here make a comment on the possibility of the emergence of warping
in the intersecting brane model introduced in the previous section. 

The warped geometry was first proposed in Ref.~\cite{rs1}, where the authors
began with a five dimensional model $M^4\times S^1/Z_2$, with
cosmological constants in the 5D bulk space and on the 4D 
boundaries, and found a solution of the Einstein equation 
which shows that the 4D metric depends on the extra dimensional
coordinate $\phi$: $ds^2 = e^{-2kr_c|\phi|}\eta_{\mu\nu}dx^\mu dx^\nu +
r^2_c d\phi^2$, where $k$ is a five dimensional scale and $r_c$ is the
compactification radius. 

It must be noted that the warping comes from localization of the
energy density.
In our setup, the internal manifold is $T^2\times T^2\times T^2$ and
orientifold planes are located at $x^4 = 0,\,\pi;\ x^6 = 0,\,\pi;\ x^8
= 0,\,\pi$, and D-brane stacks are wrapped on three cycles on the tori.
For simplifying the situation,
let us smear the D-branes and orientifold planes on the first and
second torus, not smearing them on the third torus, and assume that
the D-brane stacks projected on the third torus are all parallel to
the orientifolds. Looking at this configuration neglecting all the
extra dimensions except for $x^8$, one could regard it as the 5
dimensional model compactified on $S^1/Z_2$ with two negative tension
branes and several positive tension branes. 
Because of the localization of the positive energy branes and
the negative energy branes, it seems valid to assume the
emergence of a warp factor depending on $x_8$ though it might be small.

In the present paper, we do not attempt to solve the metric
background, but it is interesting to note that 
the authors in Ref.~\cite{Bergshoeff:2001pv} investigated a similar setup. 
They consider 10 dimensional Type IIA string theory compactified on
$S^1/Z_2$ with parallel 32 D8-branes and O8-branes. Assuming that $2n$
D8-branes are located on the top of the orientifold plane at $x_9 = 0$, and 
$32-2n$ D8-branes on the orientifold plane at $x_9 = \pi R$, where $R$
is the radius of $S^1/Z_2$, they found the solution
  \begin{eqnarray}
    \label{eq:10warping}
    ds^2 &=& H^{-1/2}_{\rm D8}[-dt^2 + (dx^\mu)^2] + H^{1/2}_{\rm D8}(dx^9)^2,
  \end{eqnarray}
where  $H_{\rm D8} = 1 - h_{\rm D8}|x^9|,\; h_{\rm D8} =
\frac{(n-8)g_s}{2\pi l_s}$, $g_s$ is the string coupling constant, and
$l_s = \sqrt{\alpha'}$ is the string length.
If we compactify this configuration on tori and T-dualize it twice, it
becomes close to our model.
\footnote{
The authors in Ref.~\cite{Chan:2000ms} suggest that 
when one would like to realize a five dimensional Randall-Sundrum type
solution from higher dimensional theory such as supergravity,
M-theory or F-theory, it would be more practical to first solve the Einstein
equations and obtain a Randall-Sundrum type solution in higher
dimensional spacetime and then perform some dimensional reduction than
to do reversely.
}
Although the procedure of torus compactification might have an
influence on the warp factor in (\ref{eq:10warping}), it is not so
audacious to expect in our setup some kind of warping along one of the
internal coordinates.

Now let us move to the evaluation of Yukawa matrices and calculate the
CKM mixing matrix.
Turning on the warp effect, we have Yukawa matrices for the up-sector:
\begin{eqnarray}
\label{eq:yukawa}
  Y^{U}_{ij} &=& e^{-k A^{U}_{ij}},\\
  A^{U}_{ij} &=& A^Q_i + A^U_j(1 - x\, c_i).
\end{eqnarray}
In this expression, we put $\epsilon_i$ in the form $\epsilon_i = x
 c_i$, where $x \ll 1$ and $c_i$ is of order 1 since the order of
 $\epsilon_i$ is thought to be of the same order.
In order to calculate the masses and the CKM matrix on the basis of
series expansion, besides $x$, we define two more small
quantities $y$ and $z$ as
\begin{eqnarray}
  y = \frac{e^{-kA_2}}{e^{-kA_3}},\quad   z = \frac{e^{-kA_1}}{e^{-kA_2}}, 
\quad(A_1 > A_2 > A_3).
\end{eqnarray}
Solving the eigensystem of Yukawa matrices by series expansion with
respect to $x$, $y$ and $z$, 
we finally obtain the leading part of up- and down- quark masses
\begin{eqnarray}
  m_{U/D,1}^2 &=& \frac{2}{9} \beta^{U/D}_4 \left\{(c^{U/D}_3)^4 + (c^{U/D}_2)^4 y^2\right\} x^4,\\
  m_{U/D,2}^2 &=& \beta^{U/D}_0 \alpha^{U/D} (c^{U/D}_2 - c^{U/D}_3)^2 y^2 x^2,\\
  m_{U/D,3}^2 &=& \beta^{U/D}_0(1 + y^2),
\end{eqnarray}
and the CKM mixing matrix 
 \begin{eqnarray}
V_{\rm CKM} &\sim& \left(\matrix{
 1 +  \frac{{n_4}\,\left( {n_2} + {n_3}\,x \right)}
     {2\,\alpha^U\alpha^D\,{\Delta^D_{23}}\,{{\Delta^{U}_{23}}}}\,z^2 
& n_4 z - \frac{{n_3}}{2\alpha^U\alpha^D\Delta^D_{23}\,\Delta^U_{23}}\,xz
& -\frac{\beta^D_1\,{n_1}}{\Delta^U_{23}}\,x\,y\,z\cr
-n_4 z+ \frac{{n_3}}
         {2 \alpha^U\alpha^D\Delta^D_{23}\,\Delta^U_{23}}\,xz  &
1 + \frac{{n_4}\,\left( {n_2} + {n_3}\,x \right)}
     {2\,\alpha^U\alpha^D\,{\Delta^D_{23}}\,{\Delta^U_{23}}}\,z^2 &
-{n_5}\,x\,y  \cr
\frac{\beta^U_1\,{n_1}}{{\Delta^D_{23}}}\,x\,y\,z & {n_5}\,x\,y & 1\cr
  }\right),
 \end{eqnarray}
where we define 
\begin{eqnarray}
n_1 &=& c^D_3\Delta^U_{12} + c^D_1\Delta^U_{23} + c^D_2\Delta^U_{31},\\
n_2 &=& -\alpha^U\alpha^D (\Delta^D_{31}\Delta^U_{23} - \Delta^D_{23}\Delta^U_{31}),\\
n_3 &=& (\beta^D_3-\beta^D_1 \beta^D_2)\alpha^U \Delta^D_{12}\Delta^D_{31}\Delta^U_{23} - \alpha^D (\beta^U_3 - \beta^U_1 \beta^U_2)\Delta^D_{23}\Delta^U_{12}\Delta^U_{31},\\
n_4 &=& \frac{\Delta^D_{31} \Delta^U_{23} - \Delta^D_{23} \Delta^U_{31}}
{\Delta^D_{23}\,\Delta^U_{23}},\\
n_5 &=& -\beta^D_1\Delta^D_{23} + \beta^U_1\Delta^U_{23},\\
\Delta^{U}_{ij} &=& c^{U}_i-c^{U}_j,
\qquad \Delta^{D}_{ij}\;=\; c^{D}_i-c^{D}_j,\\
\alpha^U &=& \beta^U_2-(\,\beta^U_1\,)^2 , \qquad
\alpha^D \;=\;\beta^D_2 - (\,\beta^D_1\,)^2,\\
\beta^U_0 &=& \sum_{j=1}^3 (\,b^{U}_j\,)^2, \qquad 
\beta^D_0 \;=\; \sum_{j=1}^3 (\,b^{D}_j\,)^2,\\
\beta^U_n &=& \sum_{i=1}^3 (\,b^{U}_i\,)^2 \left(\log b^U_i\right)^n/\,\beta^U_0, \qquad
\beta^D_n \;=\; \sum_{i=1}^3 (\,b^{D}_i\,)^2 \left(\log b^D_i\right)^n/\,\beta^D_0\nonumber\\
&&\hspace{7.5cm}\mbox{(for $n = 1, 2, \cdots$.)}
\end{eqnarray}
If we ignore coefficients and assume $x$, $y$ and $z$ of order $\epsilon$, 
we can reproduce the hierarchical structure of the realistic CKM matrix:
\begin{eqnarray}
V_{\rm CKM} &\sim& 
\left(\matrix{1 - \epsilon^2 & \epsilon & \epsilon^3\cr
\epsilon & 1 - \epsilon^2 & \epsilon^2\cr
\epsilon^3 & \epsilon^2 & 1
}\right).
\end{eqnarray}
In this na\"ive estimation, the masses of the first and the second
generations, $m_1\sim x^2$ and $m_2\sim xy$, turn out of the same order $\epsilon^2$, and one might
regard it as an undesirable result.
However, noting that $x$ comes from a small warp factor, and that it
should be smaller than $y$ and $z$, 
the mass difference is expected to come from a small value of this
ratio $x/y$ and, of course, from the coefficients we have so far
neglected.

\section{Discussion}
\label{sec:discussion}
\setcounter{equation}{0}
In this paper, we propose a method for reproducing the realistic
hierarchical structure of the observed quark masses and CKM mixing
matrix on the basis of a specific intersecting brane model proposed in
Ref.~\cite{Ibanez:2001nd}.
We newly introduced a small warp factor on tori, which
systematically changes the elements of hierarchically suppressed
but singular Yukawa matrices. The modification gives two
hierarchically suppressed quark masses and that of order 1 in addition
to a definite CKM mixing matrix. Taking the order of several 
parameters appropriately, we reproduce the realistic hierarchical
structure of quark masses and a CKM mixing matrix. 
It is worth noting that both large mass hierarchy and small mixing
angles are originated from a warp factor on the internal manifold.

\subsection*{Acknowledgments}
The author is grateful to Masahiro Yamaguchi for valuable discussions and
comments. The author also thanks the Japan Society for the Promotion
of science for financial support.

\end{document}